\PassOptionsToPackage{dvipsnames,table}{xcolor}
\documentclass[sigconf, 10pt, screen]{acmart}

\usepackage{amsmath,amssymb,amsfonts}
\usepackage{graphicx}
\usepackage{textcomp}
\usepackage[dvipsnames,table]{xcolor}
\usepackage{url}
\usepackage{siunitx}
\usepackage{courier}
\usepackage{comment}
\usepackage{xspace}
\usepackage{booktabs}
\usepackage{multirow}
\usepackage{balance}
\usepackage{microtype}
\usepackage{enumitem}
\usepackage{minted}
\usepackage{algpseudocode}
\usepackage{cleveref}
\usepackage{tcolorbox}
\usepackage{subcaption}
\usepackage{listings}
\usepackage[linesnumbered,vlined,ruled]{algorithm2e}
\usepackage{setspace}
\usepackage{tikz}

\newcommand{\sys}{\textsc{CXLMemUring}\xspace}
\newcommand{\cira}{\textsc{CXLMemUring}\xspace}

\newcommand{\FIXED}[1]{\begingroup\relax\ifmmode\else\sffamily\fi\color{OliveGreen}\ignorespaces#1\ignorespaces\endgroup}
\newcommand{\region}{hoist\xspace}
\newcommand{\regions}{hoists\xspace}
\newcommand{\Region}{Hoist\xspace}
\newcommand{\Regions}{Hoists\xspace}
\lstdefinestyle{cirair}{
  basicstyle=\ttfamily\scriptsize,
  columns=fullflexible,
  breaklines=true,
  keepspaces=true,
  frame=single,
  framerule=0.3pt,
  xleftmargin=0.5em,
  xrightmargin=0.5em,
  aboveskip=0.5em,
  belowskip=0.5em
}
\lstdefinestyle{cirairnumbered}{
  style=cirair,
  basicstyle=\ttfamily\scriptsize,
  breaklines=false,
  numbers=left,
  numberstyle=\tiny,
  numbersep=3pt,
  stepnumber=1,
  frame=single,
  framerule=0.1pt,
  rulecolor=\color{black!35},
  framesep=2pt,
  xleftmargin=1.4em,
  xrightmargin=1.4em,
  framexleftmargin=1.4em,
  framexrightmargin=1.4em,
  linewidth=\dimexpr\linewidth-2.8em\relax
}

\newcommand{\sysSpeedupLow}{1.45\xspace}
\newcommand{\sysSpeedupHigh}{1.75\xspace}
\newcommand{\sysSpeedupGeomean}{1.59\xspace}

\newcommand{\syncOverhead}{1050\xspace}               
\newcommand{\gainPerStep}{150\xspace}               
\newcommand{\vortexFreqMhz}{200\xspace}

\renewcommand\footnotetextcopyrightpermission[1]{} 
\begin{document}

\title{\sys: A Hardware Software Co-design Paradigm for Asynchronous and Flexible Parallel CXL Memory Pool Access}

\begin{abstract}
CXL-attached memory gives servers a practical path to larger memory
capacities while preserving the familiar load/store programming model.  Its
cost is latency: CXL accesses are too slow for conventional CPU mechanisms to
hide reliably, especially through dependent address-generation chains, yet too
fast for traditional software mechanisms such as context switches or
interrupt-driven asynchrony to manage at individual-load granularity.  Our
measurements on a real Granite Rapids CXL platform show the impact of this
latency gap: placing GAPBS graph workloads in CXL memory slows execution by
2.44$\times$ on average relative to local DRAM, and a state-of-the-art software
prefetcher still leaves a 2.21$\times$ slowdown.

This paper presents \sys, a hardware/software co-designed system that hides CXL
latency by using larger units of software-managed asynchrony.  \sys forms
\regions: portions of the original program that combine one or more
CXL-resident memory operations with the nearby address-generation and
memory-orchestration logic needed to execute them.  The host CPU invokes these
\regions asynchronously on a near-memory accelerator deployed on a commodity CXL
Type~2 FPGA, then continues useful computation while the device runs ahead to
prepare future accesses.  A compiler identifies candidate \regions from
unmodified source programs, and an online JIT refines region boundaries and
execution parameters based on workload-specific behavior.

We implement \sys as a prototype compiler, runtime, and Vortex-based CXL-side
accelerator.  Across GAPBS, MCF, Spatter, and NAS Parallel Benchmark workloads,
\sys improves end-to-end performance by 1.45--1.75$\times$
(1.59$\times$ geometric mean) over native CXL-memory execution. 
\end{abstract}

\author{Yiwei Yang}
\affiliation{
  \institution{UC Santa Cruz}
  \country{USA}
}
\email{yyang363@ucsc.edu}

\author{Yusheng Zheng}
\affiliation{
  \institution{UC Santa Cruz}
  \country{USA}
}
\email{yzhen165@ucsc.edu}

\author{Kexin Chu}
\affiliation{
  \institution{University of Connecticut}
  \country{USA}
}
\email{kexin.chu@uconn.edu}

\author{Jianchang Su}
\affiliation{
  \institution{University of Connecticut}
  \country{USA}
}
\email{jianchang.su@uconn.edu}

\author{Jesun Firoz}
\affiliation{
  \institution{Pacific Northwest National Lab}
  \country{USA}
}
\email{jesun.firoz@pnnl.gov}

\author{Andi Quinn}
\affiliation{
  \institution{UC Santa Cruz}
  \country{USA}
}
\email{aquinn1@ucsc.edu}

\maketitle

\section{Introduction}
\label{sec:introduction}

CXL-attached memory is becoming a practical way to expand server memory
capacity without abandoning the load/store programming model.  A CXL memory
expander can appear to software as an additional NUMA node, allowing systems to
place large working sets outside socket-local DRAM while preserving ordinary
pointer-based application code~\cite{dassharma24_cxl_intro,cxl_spec_31}.  This
model is attractive for graph analytics~\cite{gapbs},
databases~\cite{balkesen2013main}, scientific
workloads~\cite{nas_benchmarks}, language-model
serving~\cite{li2023alpaserve}, and other applications whose data
footprints continue to grow faster than per-socket DRAM capacity.

The cost of this capacity is latency.  CXL memory occupies an awkward middle
ground between local DRAM and traditional remote storage or network access.  It
is much slower than an LLC hit or local DRAM access, so the CPU's hardware
latency-hiding mechanisms---out-of-order execution, miss status holding
registers, hardware prefetchers, and deep pipelines---often cannot look far
enough ahead, especially when each address depends on the result of a previous
remote load.  At the same time, CXL latency is far shorter than the latency for
which conventional software-managed asynchrony was designed.  Threads,
coroutines, interrupts, and context switches are too expensive to deploy around
individual CXL loads.  CXL memory therefore falls into a latency ``uncanny
valley''\cite{barroso2017killer}: hardware mechanisms are too narrow to hide the latency, but
traditional software mechanisms are too coarse. 

Most existing CXL software stacks respond to this problem by improving
placement.  Page placement~\cite{maruf2023tpp}, page
migration~\cite{xiang2024nomad}, and tier
management~\cite{zhong2024memstrata} systems try to keep hot pages in
local DRAM and move colder pages to CXL memory.  These mechanisms are valuable, but they
optimize where data resides rather than what happens when a critical access to
CXL memory remains.  Host-side prefetching attacks the same latency from a
different direction, but it remains constrained by the host program's exposed
lookahead and by dependent address-generation chains~\cite{jamilan2022apt}.
For many applications, the bottleneck is not that all data is poorly placed; it is that the next useful address is discovered only after the current CXL miss returns.

Our measurements show that this problem is both large and structurally diverse.
On a real Granite Rapids CXL platform, placing GAPBS graph workloads in CXL
memory slows execution by an average of 2.44$\times$ relative to local DRAM.
Adding a state-of-the-art software prefetcher, APT-GET~\cite{jamilan2022apt}, reduces this only slightly, leaving a
2.21$\times$ slowdown.  We further study the problem to show the large number of memory-based stalls, the cost of memory indirection, and that memory boundedness varies across workloads even within an application.

This paper proposes a solution to CXL's high memory latency based upon the following key insight: rather than use an individual far-memory load as the unit for software-managed asynchrony, we should combine such loads with nearby dependencies to form regions that are large enough to amortize management overhead.  Such a region might compute the load's address and follow nearby pointer or index dependencies can be large enough to manage asynchronously. We call such a unit a \emph{\region}: a portion of the original
program that contains one or more CXL-resident memory operations and the
address-generation or memory-orchestration logic needed to execute them.


We design a new system, \sys{}, that provides a solution to high CXL memory latency by exploiting \regions{}.  \sys{} is a hardware-software co-designed solution that uses software managed asychrony to hide high CXL memory latency.  \sys{} deploys a near-memory execution accelerator alongside CXL-attached memory which computes the \regions{} that the system invokes asynchronously.  The system provides a runtime API that allows the host CPU and hardware accelerator to coordinate and synchronize for correct behavior.  This split execution model turns isolated CXL stalls into overlapped
host/device work: while the host consumes one prepared value or performs
non-memory-bound computation, the device runs ahead on address generation and
cache preparation for future accesses.

\sys is guided by two design goals.  First, \sys aims for transparency: a
developer should be able to compile an unmodified application and let the
system identify profitable \regions.  This is essential because the motivating
workloads expose many different bottleneck shapes, and manually rewriting each
application around a new device programming model would defeat the purpose of
using CXL as coherent memory.   To support transparency, \sys{} uses a compiler-based approach to split an unmodified source program into regions.  Deciding on the appropriate \regions{} is difficult: overly small \regions{} struggle to hide management overhead while overly large \regions{} require too much computation to happen on a relatively weak hardware accelerator.  \sys{} uses a data-flow based algorithm inspired by program slicing to create initial \regions{} and uses an online just-in-time compiler to refine and optimize its \region{} decisions based upon workload-specific behavior. 

\sys{}'s second design goal is commodity support, meaning that \sys{} should work without requiring custom ASICs or modifications to the host CPU.  To ensure commodity support, \sys{}'s hardware accelerator is deployed on an existing CXL Type~2 device, namely a CXL Agilex 7 FPGA, rather than on a custom ASIC.  In addition, \sys{}'s host-side implementation is entirely written as a software layer that requires no special features from the host CPU.

We implement \sys as a prototype compiler and runtime stack.  The compiler adds
approximately 12,000 lines of C++ to MLIR.  It supports programs written and compiled through Clang~\cite{mlir_llvm}, TOSA~\cite{mlir_tosa_dialect}, and SCF~\cite{mlir_scf}.  The system implements an eight-thread Vortex SIMT cores as its hardware accelerator ~\cite{clangir23}.  Its runtime implements the system's coordination and synchronization using a CXL-backed shared queue.

We evaluate \sys{} on benchmarks from GAPBS~\cite{gapbs}, MCF, Spatter~\cite{spatter_memsys20}, and the NASA Parallel Benchmark~\cite{nas_benchmarks} workloads.  We observe that
\sys improves end-to-end performance by 1.45--1.75$\times$
(1.59$\times$ geometric mean) over native CXL memory performance.  We also analyze \sys{}'s desgin decisions and compare alternative policies for deciding how to form \regions{} from a program.

In summary, this paper makes the following contributions:

\begin{itemize}[leftmargin=*]
    \item \regions as a unit of software-managed asynchrony that are large enough to amortize
    synchronization overhead.

    \item \sys, a transparent and commodity-supported hardware/software codesigned system that uses software managed asychrony to hide CXL memory latency.
    
    \item An evaluation of \sys that shows that it improves application performance on CXL-memory machines by up to a factor of \sysSpeedupHigh.
\end{itemize}

\section{Background and Motivation}
\label{sec:motivation}

CXL-attached memory increases memory capacity and enables new forms of
disaggregated memory, but it also exposes applications to remote-memory
latency that is not hidden well by existing page-placement or prefetching
mechanisms.

\subsection{CXL Primer}

Compute Express Link (CXL) is an open interconnect built on top of PCIe that gives processors a cache-coherent path to attached devices and device memory regions~\cite{dassharma24_cxl_intro,cxl_spec_31}. CXL supports memory expansion through Type~3 devices, which expose device memory to the host.  Operating systems manage the CXL memory and expose it as a NUMA node to applications, which allows applications to allocate large working sets to the CXL memory.  Applications use the same load/store programming model to access CXL memory as it uses for local DRAM.  This effectively allows software to grow its memory capacity beyond socket-local memory capacity without requiring software modifications that are often required to support non-socket memory (e.g., RDMA, I/O interfaces, etc.). 

CXL also defines interfaces that allow devices to participate in the host CPU's memory coherency through Type~2 devices.  These devices can issue device-to-host coherency requests through a device coherence agent.  

CXL's cost is latency. A CXL memory access uses ordinary load/store semantics, but a miss to CXL-attached memory must traverse the CXL link and device path before the host can consume the data. This latency is much larger than a local DRAM access.  


\subsection{Evaluating systems with CXL Memory}

We first measure how applications perform when their data is placed in CXL memory. Our goal is to quantify the application impact of CXL placement and to test whether an existing host-side latency-hiding technique is sufficient. 

\paragraph{Experimental Setup}  We use a real CXL platform with an Intel Xeon~6 6787P host, 256\,GB of local DDR5-8000, and CXL Agilex 7 FPGA with 16GB of DDR4 and a delay buffer imposing a memory latency of 1us.  The CXL module is enumerated by platform firmware as a CPU-less NUMA node. For the CXL
configuration, benchmarks are launched with \texttt{numactl} and bound to the CXL node, so their allocations reside on the CXL module while CPU placement is held fixed. We use the graph workloads from the GAP Benchmark Suite~\cite{beamer2015gap} because their irregular traversals and dependent updates are a common stress case for far-memory systems; we configure the experiments to use a graph with $2^{20}$ vertices.  We compare three configurations. \emph{DRAM} places workload data in local DDR memory. \emph{CXL} places the data in the CXL memory node. \emph{CXL+ APT-GET} keeps data in CXL memory and adds APT-GET~\cite{jamilan2022apt} host prefetch insertion, representing an existing state-of-the-art approach for hiding memory latency from the CPU side.

\begin{figure}[t]
\centering
\includegraphics[width=\linewidth]{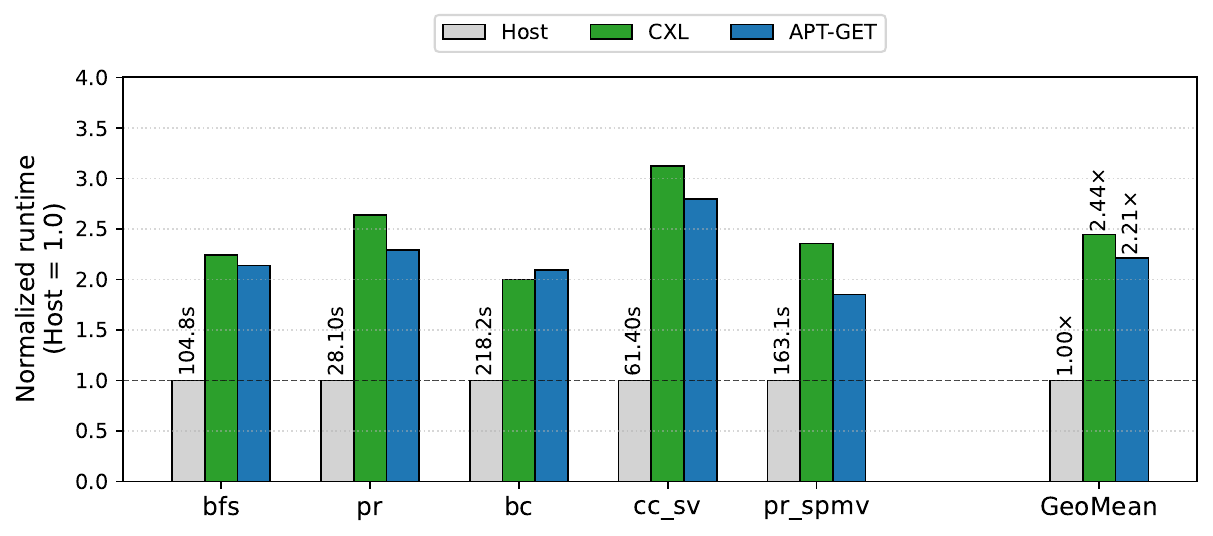}
\caption{GAPBS performance on baseline systems}
\label{fig:motivation:experiment}
\end{figure}

\paragraph{Results} We measure the latency of each benchmark under each of the configurations and plot the results in \cref{fig:motivation:experiment}.  Across the GAPBS kernels, moving data from local DRAM to CXL memory imposes a slowdown of an average of 2.44$\times$.  The core issue is that the graph processing workloads' irregular graph traversals repeatedly place remote CXL accesses on the critical path.  

APT-GET's host prefetching does not make much impact on this gap; APT-GET retains a slowdown of 2.21$\times$ compared to using local DRAM.  The core issue is that APT-GET's software prefetching approach cannot prefetch far enough ahead to hide the high CXL memory latency because the pointer-chasing operations in the benchmarks are too tightly packed for software prefetching to be effective.

\subsection{Understanding Bottlenecks When using CXL Memory}
\label{sec:vtune}

We next study why CXL-resident applications slow down.

\paragraph{Experimental Setup} This analysis uses the same Granite Rapids CXL platform as above and adds two analysis tools. First, we use Intel VTune Top-Down Microarchitecture Analysis (TMA) to measure where CPU pipeline slots are lost when allocations are bound to the CXL memory node. TMA classifies issue slots into frontend, bad speculation, retiring, and backend categories; the backend category is further split into core-bound and memory-bound behavior. Second, we use a cycle-level gem5 model to isolate pointer-chasing cost under controlled access patterns. The gem5 study lets us hold the remote-memory path fixed while changing only the amount of address metadata that must be fetched before a useful access can issue. For Vtune, we use selected benchmarks from 3 benchmark suites:  is and gc from the NAS Parallel Benchmarks~\cite{nas_benchmarks}; bc and pr from GAPBS~\cite{beamer2015gap}; partitioned hash-join kernels~\cite{partitioned_hash_join}; GZP and GZPI from the UME benchmark suite~\cite{henning2023ume}; and Spatter gather/scatter accesses~\cite{spatter_memsys20}.  Together these workloads cover sequential streams, single indirection, double indirection, read-modify-write updates, and conditional irregular accesses.

\begin{table}[t]
\caption{TMA Top-Down Microarchitecture Summary. All percentages are of
pipeline slots. }
\label{tab:tma_summary}
\resizebox{\columnwidth}{!}{%
\begin{tabular}{@{}llrrr@{}}
\toprule
\textbf{Benchmark} & \shortstack{\textbf{Hotspot}\\\textbf{Instruction}} & \shortstack{\textbf{Backend}\\\textbf{Bound}} & \shortstack{\textbf{Memory}\\\textbf{Bound}} & \textbf{Verdict} \\ \midrule
is & RMW A[B[i]] & 78.1\% & 64.5\% & Memory \\
gc & LD A[B[j]] & 59.9\% & 39.5\% & Memory \\
bc & RMW A[B[j]] cond & 74.2\% & 62.2\% & Memory \\
pr & RMW A[B[j]] & 78.7\% & 65.0\% & Memory \\
Hash-Join PRH & ST A[B[f(C[i])]] & 57.4\% & 38.6\% & Memory \\
Hash-Join PRO & ST A[B[f(C[i])]] & 63.5\% & 42.2\% & Memory \\
UME GZP & RMW A[B[i]] cond & 70.6\% & 33.9\% & Mem+Core \\
UME GZPI & LD A[B[C[j]]] cond & 69.9\% & 41.7\% & Memory \\
Spatter & ST A[B[i]] & 72.8\% & 31.1\% & Mem+Core \\
\bottomrule
\end{tabular}%
}
\end{table}

\paragraph{Results} We provide the key results from top down analysis in \Cref{tab:tma_summary}.  Every listed kernel loses at least 30\% of pipeline slots to memory stalls, while is, bc, and pr each lose more than 60\%. The affected code is not limited to one benchmark family: graph analytics, sparse numerical kernels, hash joins, unstructured meshes, and gather/scatter kernels all expose substantial memory stalls.  VTune reports that memory is the key bottleneck for all applications but two, where it reports that both Core and Memory lead to many stalls.  Moreover, VTune reports indirect memory accesses as the hotspot instruction, i.e., the instruction that accounts for the most overhead, for each of the benchmarks.

\begin{table}[t]
\caption{Throughput results of microbenchmarks in a gem5 simulator}
\label{tab:cxl_addrgen_motivation}
\footnotesize
\centering
\begin{tabular}{@{}lrr@{}}
\hline
\textbf{Mode} &
\shortstack{\textbf{Throughput}\\\textbf{(instructions/us)}} &
\shortstack{\textbf{Slowdown}\\\textbf{vs. direct}} \\
\hline
direct load (\texttt{*ptr})   & 0.50 & 1.00$\times$\\
indirect load (\texttt{*A[i]})    & 0.26 & 1.91$\times$\\
doubly-indirect load (\texttt{*A[B[i]]}) & 0.17 & 2.94$\times$\\
\hline
\end{tabular}%
\end{table}

We further explore the cost of irregular memory accesses on CXL memory systems by using the gem5 simulator.  We implement microbenchmarks that repeatedly issue direct load instructions, indirect instructions, and doubly indirect instructions.  \Cref{tab:cxl_addrgen_motivation} shows the throughput of each microbenchmark as well as the throughput slowdown relative to direct loads.  We observe that adding a layer of indirection---i.e., a single indirect load compared to a direct load---halves the throughput of the microbenchmark; a doubly indirect load adds similar overhead. 

\begin{table}[t]
\caption{MonetDB TPC-H per-query hotspot analysis at SF=1 on Granite Rapids.
Percentages are of total CPU time (kernel spinlock cycles from 172-thread
contention excluded).}
\label{tab:monetdb_hotspots}
\resizebox{\columnwidth}{!}{%
\begin{tabular}{@{}llll@{}}
\toprule
\textbf{Query} & \textbf{Function 1} & \textbf{Function 2} & \textbf{Function 3} \\ \midrule
Q1 (Agg) & \texttt{BATgroupavg3} (25.0\%) & \texttt{dosum} (11.2\%) & \texttt{project1\_lng} (9.5\%) \\
Q5 (Join+Filter) & \texttt{hashjoin} (6.9\%) & \texttt{scanselect} (3.2\%) & \texttt{mergejoin\_lng} (2.8\%) \\
Q6 (Scan/Filter) & \texttt{scanselect} (17.1\%) & \texttt{canditer\_next} (1.5\%) & -- \\
Q9 (Join+Agg) & \texttt{hashjoin} (31.4\%) & \texttt{canditer\_next} (3.9\%) & -- \\
Q11 (Join+Agg) & \texttt{hashjoin} (58.9\%) & \texttt{canditer\_next} (8.0\%) & \texttt{thetajoin} (1.6\%) \\
Q13 (String+Group) & \texttt{BATgroup\_internal} (16.2\%) & \texttt{\_\_strstr} (15.7\%) & \texttt{PCRElikeselect} (5.3\%) \\
Q16 (Group+Sort) & \texttt{BATgroup\_internal} (16.9\%) & \texttt{GDKqsort} (4.4\%) & \texttt{\_\_strcmp\_evex} (5.3\%) \\
Q18 (Join+Agg) & \texttt{dosum} (18.1\%) & \texttt{BATgroup\_internal} (14.1\%) & -- \\
Q22 (Join+String) & \texttt{hashjoin} (53.8\%) & \texttt{UTF8\_strtail} (5.0\%) & \texttt{scanselect} (3.7\%) \\
\bottomrule
\end{tabular}%
}
\end{table}

Lastly, we explore the variance in memory bottlenecks when we vary the workloads for a given application.  We randomly select candidate queries from TPC-H and execute them over MonetDB~\cite{monetdb} deployed on the Granite Rapids CXL platform.  We configure the experiment to use scale factor 1.  We use VTune to determine the most expensive functions when running each of the queries; \cref{tab:monetdb_hotspots} reports the top three functions for each query (it only shows two if the third function represents less than 1\% of execution time).  We observe that the dominant function varies considerably accross queries, suggesting that memory bottlenecks are workload dependent.
\begin{figure}[t]
     \centering
    \resizebox{0.98\linewidth}{!}{
\usetikzlibrary{positioning, fit, arrows.meta, backgrounds, calc}

\begin{tikzpicture}[
  font=\small,
  >={Stealth[scale=0.65]},
  every node/.style={align=center},
  hbox/.style={draw=blue!60!black, fill=blue!9, rounded corners=3pt,
               minimum width=2.1cm, minimum height=0.65cm, inner sep=2pt},
  dbox/.style={draw=teal!65!black, fill=teal!9, rounded corners=3pt,
               minimum width=3.3cm, minimum height=0.65cm, inner sep=2pt},
  cbox/.style={draw=orange!70!black, fill=orange!10, rounded corners=3pt,
               minimum width=2.4cm, minimum height=0.65cm, inner sep=3pt},
  grpbox/.style={draw=gray!55, dashed, rounded corners=5pt, inner sep=3pt, thick},
  lbl/.style={font=\scriptsize\bfseries, anchor=south west, inner sep=0pt},
  farr/.style={->, thick},
  barr/.style={<->, thick},
]


\node[cbox] (src)  at (4.5, 6.6) {Source Program};
\node[cbox] (comp) at (4.5, 5.9) {\textbf{CIRA Compiler}};
\draw[farr] (src) -- (comp);

\node[minimum size=0pt, inner sep=0pt] (ct_bpad)
  at ([yshift=-0.35cm]comp.south) {};

\begin{scope}[on background layer]
  \node[grpbox, fit=(ct_bpad)(src)(comp)] (ctbox) {};
\end{scope}
\node[lbl] at ([xshift=3pt, yshift=3pt]ctbox.south west) {Compile Time};


\node[hbox] (hbin) at (1.6, 4.2) {Host Binary};
\node[hbox] (jit)  at (1.6, 3.2) {CIRA JIT};
\node[hbox] (hrt)  at (1.6, 2.2) {Host Runtime};
\node[hbox] (hcpu) at (1.6, 1.2) {Host CPU};

\draw[barr] (hbin) -- (jit);
\draw[barr] (jit)  -- (hrt);
\draw[barr] (hrt)  -- (hcpu);

\node[minimum size=0pt, inner sep=0pt] (h_bpad)
  at ([yshift=-0.35cm]hcpu.south) {};

\begin{scope}[on background layer]
  \node[grpbox, fit=(h_bpad)(hbin)(jit)(hrt)(hcpu)] (hgrp) {};
\end{scope}
\node[lbl] at ([xshift=3pt, yshift=3pt]hgrp.south west) {Host System};


\node[dbox, minimum width=1.6cm] (cmem) at (4.4, 1.2) {CXL\\ Memory};

\node[dbox] (fpga) at (7.0, 1.2) {CXL-attached\\ FPGA};
\node[dbox] (vacc) at (7.0, 2.2)
  {Execution Accelerator\\[1pt]{\scriptsize (Vortex RISC-V SIMT)}};
\node[dbox] (drt)  at (7.0, 3.2) {Device Runtime};
\node[dbox] (rbin) at (7.0, 4.2) {Region Binary};

\draw[barr] (cmem.east) -- (fpga.west);
\draw[barr] (fpga) -- (vacc);
\draw[barr] (vacc) -- (drt);
\draw[barr] (drt)  -- (rbin);

\node[minimum size=0pt, inner sep=0pt] (d_bpad)
  at ([yshift=-0.35cm]fpga.south) {};

\begin{scope}[on background layer]
  \node[grpbox, fit=(d_bpad)(cmem)(fpga)(vacc)(drt)(rbin)] (dgrp) {};
\end{scope}
\node[lbl] at ([xshift=3pt, yshift=3pt]dgrp.south west) {Device};


\node[minimum size=0pt, inner sep=0pt] (rt_bpad_h)
  at ([yshift=-0.35cm]hgrp.south) {};
\node[minimum size=0pt, inner sep=0pt] (rt_bpad_d)
  at ([yshift=-0.35cm]dgrp.south) {};

\begin{scope}[on background layer]
  \node[grpbox, fit=(rt_bpad_h)(rt_bpad_d)(hgrp)(dgrp)] (rtbox) {};
\end{scope}
\node[lbl] at ([xshift=3pt, yshift=3pt]rtbox.south west) {Runtime};


\draw[barr, draw=teal!65!black]
  (hcpu.east) --
  node[above, font=\scriptsize, color=black] {CXL}
  (cmem.west);


\draw[farr] (comp.south) -- ++(0,-0.08) -| (hbin.north);
\draw[farr] (comp.south) -- ++(0,-0.08) -| (rbin.north);

\end{tikzpicture}}
     \caption{\sys{}'s architecture.}
     \label{fig:cira_arch}
\end{figure}
\section{CIRA Design}
\label{sec:design}

\sys is a new system for software-managed asynchrony for hiding CXL memory latency. \sys's key insight is that the benefits of software-managed asynchrony---i.e., support for larger windows of asynchronous operations---are necessary to hide memory latency at the magnitude of CXL memory, while the costs of software-managed asynchrony---i.e., high overhead---are manageable by performing groups of operations asynchronously instead of a single load operation.  \sys calls such groups of operations a \emph{\region} (\cref{sec:design:api}).  \sys enables \regions by deploying a near-memory execution accelerator alongside CXL-attached memory (\cref{sec:design:arch}).  Choosing appropriate \regions from a program is challenging; \sys uses a multi-staged approach including a custom compiler and dynamic optimizer to optimize \region selection (\cref{sec:design:compiler} and \cref{sec:design:jit}).

\sys's design is guided by two design goals. First, \sys aims for
\emph{transparency}, meaning that a developer can use \sys without
modifying the application. Transparency is a standard goal in prior
systems for hiding high-latency memory and storage operations,
including far-memory runtimes, asynchronous prefetching systems,
and transparent memory-tiering frameworks~\cite{ruan2020aifm,maruf2020leap,jamilan2022aptget,maruf2023tpp}, so \sys retains this goal.
Second, \sys aims for \emph{commodity support}, meaning that
\sys should work without requiring custom ASICs or modifications
to the host CPU.

\subsection{System Architecture}
\label{sec:design:arch}

\Cref{fig:cira_arch} shows the \sys architecture.  The system comprises four main components: the \sys hardware consists of a near-memory execution accelerator designed to run on commodity CXL Type~2 FPGA devices deployed alongside CXL-attached memory; its approach ensures commodity support.  The \sys runtime consists of software that executes on both the accelerator and the CPU host to implement the \cira API and coordinate communication between the two sides.  The \sys compiler is responsible for compiling a program to split it into \regions and use \sys's runtime system to accelerate CXL-attached memory accesses.  Finally, the \sys JIT dynamically optimizes a running process to maximize the benefit of asynchronous actions, adapting \region selection and parameters to observed runtime behavior.

The remainder of this section describes \sys{}'s key components.  \Cref{fig:design:example} provides an example program that uses the \sys API.  The original program (\cref{fig:design:example:orig}) is a linked-list traversal that calls a function, \texttt{foo}, on every element of the linked list.  \sys converts the program into two separate components, one that executes on the host CPU (\cref{fig:design:example:host}) and one consisting of a \region that executes on the device (\cref{fig:design:example:compiled}).

\begin{figure}[t]
\centering
\begin{list}{}{\leftmargin=2.4em\rightmargin=-2.4em\topsep=0pt\partopsep=0pt\parsep=0pt\itemsep=0pt}
\item\relax
\begin{minipage}{0.97\linewidth}
\begin{lstlisting}[style=cirairnumbered,basicstyle=\ttfamily\footnotesize,escapeinside={(*@}{@*)}]
Node *node = head;
while (node) {
  foo(node->data);
  node = node->next;
}\end{lstlisting}
\subcaption{Original Program.}
\label{fig:design:example:orig}
\end{minipage}

\begin{minipage}{0.97\linewidth}
\begin{lstlisting}[style=cirairnumbered,basicstyle=\ttfamily\footnotesize,escapeinside={(*@}{@*)}]
host_side(head) {
  buf = cira_alloc(sizeof(Value));(*@\label{fig:design:example:alloc}@*)
  db  = cira_doorbell_alloc();
  node = head;
  cira_hoist_exec(hoist, node, buf, db);(*@\label{fig:design:example:hoist_call}@*)
  while (node) {
    cira_doorbell_wait(db);(*@\label{fig:design:example:host_wait}@*)
    foo(*buf);
    cira_doorbell_ring(db);(*@\label{fig:design:example:host_ring}@*)
  }
  cira_free(buf);(*@\label{fig:design:example:free}@*)
} 
\end{lstlisting}
\subcaption{Host portion of the traversal using \sys API}
\label{fig:design:example:host}
\end{minipage}

\begin{minipage}{0.97\linewidth}
\begin{lstlisting}[style=cirairnumbered,basicstyle=\ttfamily\footnotesize,escapeinside={(*@}{@*)}]
hoist(host, buf, db) {
  while (node) {
    *buf = node->data;
    node = node->next;
    cira_doorbell_ring(db);(*@\label{fig:design:example:device_ring}@*)
    cira_doorbell_wait(db);(*@\label{fig:design:example:device_wait}@*)
  }
}
\end{lstlisting}
\subcaption{\region for the traversal using \sys API}
\label{fig:design:example:compiled}
\end{minipage}
\end{list}
\caption{An example linked list program.}
\label{fig:design:example}
\end{figure}

\subsection{The CIRA API}
\label{sec:design:api}

\begin{table}[t]
\caption{CIRA API}
\label{tab:cira_ops}
\centering
\footnotesize
\begin{tabular}{@{}p{0.48\linewidth}@{\hspace{0.6em}}p{0.45\linewidth}@{}}
\toprule
\textbf{Operation} & \textbf{Description} \\
\midrule
\texttt{cira\_hoist\_exec(fxn, ...)} & Execute \texttt{fxn} on the device, passing all optional arguments. \\
\texttt{p = cira\_alloc(size)} & Allocate a buffer of \texttt{size} bytes in CXL memory. \\
\texttt{cira\_free(p)} & Free pointer, \texttt{p}, from CXL memory.\\
\texttt{d = cira\_doorbell\_alloc()} & Allocate a doorbell for synchronization between host and device.\\
\texttt{cira\_doorbell\_wait(d)} & Pause execution until the other side rings the doorbell. \\
\texttt{cira\_doorbell\_ring(d)} & Wake the other side of the doorbell. \\
\bottomrule
\end{tabular}%
\end{table}

We describe the key features and components of the \sys{} API; \cref{tab:cira_ops} provides a summary of the operations that make up the API.

\paragraph{\Region} \sys's API is to support \regions. A \region is a portion of the original program that \sys offloads and executes on its execution accelerator. \Regions are pure functions (i.e., side-effect free) and cannot cross critical sections.  Otherwise, they can include arbitrary code from the original application. The host invokes a \region by executing the \texttt{cira\_hoist\_exec} function with a pointer to the region and any optional arguments.  For example, line~\ref{fig:design:example:hoist_call} in \cref{fig:design:example:host} shows how the linked list host program initiates the \region for the linked list traversal example.

\paragraph{Memory management} \sys{} provides APIs for CXL-memory allocation through \texttt{cira\_alloc} and release through \texttt{cira\_free}.  In addition to using these APIs to store application data, \sys{}'s memory management provides a mechanism to create objects that are visible to both the host and device~\footnote{\sys{}'s hardware setup does not allow the device to access the host's DRAM.}. The programming model is simple: the host allocates a buffer for shared data and passes the pointer of the buffer to the \region that it wishes to share with.  For example, lines~\ref{fig:design:example:alloc} and \ref{fig:design:example:free}  in \cref{fig:design:example:host} show how the example host program uses these functions to allocate and free a buffer for communicating values to and from the device.

\paragraph{Doorbells} \sys{} provides a \texttt{doorbell} API to support synchronization between the host and device. A program allocates a doorbell with the API call in Table~\ref{tab:cira_ops}. Calling \texttt{cira\_doorbell\_wait} causes the calling context (host or device) to pause execution, while calling \texttt{cira\_doorbell\_ring} from one context wakes up the other. Rings to a doorbell are stateful: if the other context is not currently waiting, its \emph{next} wait will not pause. Rings are also coalesced: two rings without an intervening wait have the same effect as one ring. For example, the linked-list host and device programs use a doorbell (\texttt{db}) to safely synchronize accesses to the shared \texttt{buf} variable: the host waits on line~\ref{fig:design:example:host_wait} before consuming \texttt{buf} and rings the doorbell on line~\ref{fig:design:example:host_ring} after it is done, while the device rings on line~\ref{fig:design:example:device_ring} after producing \texttt{buf} and waits on line~\ref{fig:design:example:device_wait} before overwriting it.

\subsection{CIRA Hardware}
\label{sec:design:hardware}



\sys deploys a near-memory execution accelerator to execute \regions. The system uses commodity FPGA hardware for the accelerator, implemented on an Intel Agilex~7 FPGA~\cite{ji2024demystifying}. Rather than modifying the host CPU~\cite{wang2024asynchronous,talati2021prodigy} or introducing a custom host-device interface~\cite{dx100,kuper2024quantitative,berthold2024demystifying}, \sys uses CXL as the communication substrate. In particular, the FPGA is exposed as a Type~2 CXL device, allowing the accelerator to participate in the host coherence domain and access CXL-resident data through standard coherent memory operations.

\sys employs Vortex RISC-V single-instruction multiple-thread (SIMT) cores~\cite{elsabbagh19_vortex} as its execution substrate. Although these cores run at a much lower clock frequency than the host CPU, this is a good match for \sys because \regions are dominated by remote-memory latency rather than scalar compute throughput. Our intuition is that the goal is not to make one dependence chain execute faster instruction-by-instruction, but to keep many independent address-generation and prefetch contexts in flight while the host performs useful work. The compiler-formed regions and profile-guided batching described above expose this concurrency: pointer-chasing and graph-traversal kernels can run ahead across multiple nodes, frontiers, or batched requests, issuing many outstanding CXL.cache accesses and switching among lightweight threads when individual accesses stall. In this setting, high SIMT concurrency converts otherwise serialized host stalls into overlapped device-side memory-level parallelism, while the low per-core clock rate is largely hidden by the memory-bound nature of the workload.

\subsection{CIRA Software Runtime}
\label{sec:cira_runtime}

\sys's software runtime implements the \cira API by coordinating a host runtime library with a small device-side service loop. 

The \sys runtime coordinates \region execution through a shared queue in CXL memory.  When host code calls \texttt{cira\_hoist\_exec}, the runtime packages the selected \region identifier and its arguments into a task descriptor and appends that descriptor to the queue.  The device runtime polls the queue, removes ready descriptors, and dispatches the corresponding \region on the Vortex cores. 

The \sys runtime provides its memory-management APIs using conventional allocator machinery over CXL-backed memory, rather than a custom object system. 

\sys implements doorbells as small shared synchronization objects. Each doorbell records whether the host or device has already rung it and whether either side is currently waiting. A call to \texttt{cira\_doorbell\_wait} polls this state until the corresponding ring is visible; a call to \texttt{cira\_doorbell\_ring} updates the state so the peer's current or next wait can proceed. The runtime inserts the necessary CXL coherency operations around these polling loads and state updates so that data written before a ring is visible to the peer after its wait returns. Thus, the doorbell API gives generated host and device code a simple synchronization abstraction while the runtime handles the low-level sharing details.

\subsection{CIRA Compiler}
\label{sec:design:compiler}

The \sys compiler converts traditional application code into coordinated host and device programs that use the \sys API to coordinate and share state.  The key task for the compiler is to determine what \regions to make from the original program.  \sys solves this task through a slicing-based program analysis that aims to balance \region size: it ensures that \regions are large enough so that they overcome the overhead of its software management.  However, \sys also must ensure not to place too much computation in \regions, since \region logic executes on the execution accelerator, which is significantly less powerful than the host CPU. 

The \sys compiler first compiles the program to produce the MLIR.  The \sys compiler first identifies all candidate \region points.  The system considers all \texttt{load}s from CXL memory that occur within a loop as candidate \region points.  Loads from CXL memory are precisely the instructions whose latency \sys aims to hide, while focusing on those that occur in loops ensures that the created \regions will be of high impact. 

Next, the \sys compiler expands each candidate load to grow the \region and amortize the overhead of software management.  The system starts at the candidate load instruction and recursively adds all operations that are necessary to compute the input to the instruction.  For example, the system first adds the operations that define the address operand to the candidate load instruction, then adds operations that define the inputs to the inputs to the address operand, and so on.  It stops the recursion and chooses to treat an operand as an input to the \region if one of the following conditions holds: 1. \sys cannot execute the candidate instruction,  2. \sys's recursion has reached the start of the function that contains the original candidate load, 3. \sys's recursion would cross a synchronization operation (i.e., adding the instruction would cause the \region to cross a critical section boundary), 4. the candidate instruction is invariant with the loop, or 5. the candidate instruction belongs to a basic block that is not currently in the \region and the branch predicate of the new basic block cannot be safely added to the \region due to any of these reasons.  As the final step in expansion, the \sys compiler adds as many other CXL loads to the \region as it can. It identifies all of the uses of each of the instructions in the \region and adds any such uses that are CXL memory loads and are not currently in the \region. 

To create the host-side, the \sys compiler attempts to remove all instructions from the \region.  Note that it may decide to leave \region instructions in the host program, thereby duplicating the instruction, if the host program includes uses of the \region instructions. 

Finally, the \sys compiler adds synchronization to both the host program and the \region to ensure that the host and device communicate data safely.  During this step, the compiler also ensures that both programs include the loop that contains the original candidate load.

As an example, consider the original linked list traversal described in \cref{fig:design:example:orig}.  The first candidate instruction is the load of the node's next pointer (i.e., \texttt{node = node->next}).  The compiler identifies the instructions that define the original value of \texttt{node}, which includes either itself (which it does not need to add) or the initialization of \texttt{node} to \texttt{head}~\footnote{\sys operates over LLVM IR, so this actually involves tracing single static assignment operations and phi nodes}. The compiler does not add the initialization operation because it is loop invariant; instead it creates an input to the region for the original \texttt{node} value.  As the last step in expanding the region, \sys{} identifies that the cxl load (\texttt{node->data}) depends on instructions that are already contained in the \region and adds it.

\subsection{CIRA JIT}
\label{sec:design:jit}

\begin{figure}[t]
\centering
\begin{minipage}{0.49\linewidth}
\begin{lstlisting}[basicstyle=\ttfamily\fontsize{5.5}{5.5}\selectfont,escapeinside={(*@}{@*)}]
host_side(head) {
  B=16; db=cira_doorbell_alloc();
  data=cira_alloc(sizeof(Value)*B);
  nodes=cira_alloc(sizeof(Node*)*B);
  nodes[0]=head; n=head;
  cira_hoist_exec(hoist,nodes,data,db);

  while (n) {
    cira_doorbell_wait(db);
    n=nodes[0];
    for (i=0; n && i<B; i++, n++)
      foo(data[i]);
    cira_doorbell_ring(db);
  }
  cira_free(data); cira_free(nodes);
}
\end{lstlisting}
\subcaption{Host}
\label{fig:design:example:host}
\end{minipage}\hfill
\begin{minipage}{0.49\linewidth}
\begin{lstlisting}[basicstyle=\ttfamily\fontsize{6}{6}\selectfont,escapeinside={(*@}{@*)}]
hoist(nodes, data, db) {
  B=16; n=nodes[0];

  while (n) {
    for (i=0; i<B && n; i++) {
      data[i]=n->data;
      nodes[i]=n;
      n=n->next;
    }
    nodes[0]=n;
    cira_doorbell_ring(db);
    cira_doorbell_wait(db);
  }
}
\end{lstlisting}
\subcaption{\region}
\label{fig:design:example:compiled}
\end{minipage}
\caption{The \sys JIT's optimized version of the linked-list traversal in \cref{fig:design:example}.}
\end{figure}

\sys implements a JIT to adapt and optimize the \regions for the particular workload that a user executes; the motivation shows how varied hotspots and optimizations can be across workloads even within an application (\cref{sec:motivation}).  The \sys JIT uses a two-phase approach: a profile phase uses performance counters to identify performance bottlenecks as the system operates.  Then, an optimization phase uses those profiles to guide a rule-based optimizer. 

\paragraph{The Profiler} The \sys JIT's profiling system uses a hierarchical approach to identify optimization opportunities in the program.  At the inner level,  the system uses hardware performance counters on both the host CPU and the hardware accelerator to track individual memory access latencies and host CPU cache behavior.  At the outer level, the system identifies program phases with distinct memory access characteristics, by monitoring the functions in which the process spends the majority of its time. 

\paragraph{The Optimizer}  The \sys optimizer uses the data from the \sys profiler to optimize the program.  It initiates an optimization either when the profiling data indicates poor performance (e.g., a low cache hit rate) or a shift in program phases (e.g., the hot functions in the program shift).  The optimizer includes hysteresis to prevent the system from over-corrections that would cause it to oscillate between configurations by incorporating an exponential backoff for optimization attempts that fail.

When the optimizer decides to optimize a \region, it has a number of options.  First, it can adjust the operations in the \region and the host program to shrink the \region from the output produced by the compiler.  It shrinks \regions if the profiling data suggests that the host is frequently stalled waiting on the device.  Second, the optimizer can expand the region by unrolling loops and using batching. Growing the batch size has the effect of growing the \region, which is useful if the profiling data suggests that the device is frequently waiting for the host.  Finally, the optimizer considers adjusting when the host initiates \region execution (i.e., calls \texttt{cira\_hoist\_exec}) to hide high startup cost. 

For example, \cref{fig:design:example:compiled} shows the linked list traversal code optimized by the \sys JIT.  In this case the \sys JIT used batching to grow the size of the original \region. 

\section{Implementation}

\subsection{Compiler Infrastructure}

The implementation of \sys required approximately 12,000 lines of C++ code added to the MLIR compiler infrastructure~\cite{mlir_llvm}, spanning frontend parsing, IR transformations, and backend code generation. \sys supports three frontend dialects: ClangIR for general C/C++ code, TOSA for tensor computations, or SCF for structured control-flow kernels~\cite{clangir23,kunze21_tosa,mlir_tosa_dialect,mlir_scf}.  The ClangIR frontend integration involved extending Clang's AST visitor patterns to generate MLIR operations while preserving C++ semantic information. We implemented custom parsers for TOSA and SCF dialects that maintain source-level debugging information through the compilation pipeline.

The frontend parsing infrastructure handles modern C++ codebases through a multi-phase translation process. During the first phase, Clang's semantic analysis produces an annotated AST with type information, template instantiations, and implicit conversions made explicit. The second phase walks this AST to generate ClangIR operations, preserving the high-level structure of classes, virtual dispatch, and exception handling. The third phase lowers ClangIR to SCF and standard dialects, converting control flow constructs to structured operations amenable to analysis. Throughout this process, we maintain bidirectional mappings between source locations and IR operations.

\subsection{Backend Code Generation}

Backend code generation required developing custom LLVM target descriptions for our heterogeneous architecture. The x86 backend extensions include 15 new intrinsics for asynchronous memory operations, completion polling, and CXL-visible queue manipulation, mapping to instruction sequences that interact with our runtime system. The Vortex backend implements a specialized instruction scheduler that accounts for CXL memory latency and policy-specific coherence costs when ordering prefetch operations and distributing work across warps. During lowering, \texttt{cira.cache\_read} and \texttt{cira.cache\_write} are translated into R-Tile request-generator commands carrying route and policy bits: CS/CO/NC for reads, NC/CO/NC-P for writes, and plain CXL.mem loads/stores for H2D accesses. We developed custom register allocators for both backends that minimize spill traffic to CXL memory.

\subsection{Runtime System}

The runtime system follows a distributed design with components executing on both x86 host processors and Vortex accelerator cores. The host-side runtime, implemented as a shared library, intercepts memory allocation requests and manages the interface to the Vortex subsystem. The accelerator-side runtime, compiled into a persistent kernel resident in Vortex memory, processes state-aware prefetch and publication requests, records the most recent compiler-requested state for outstanding cache lines, and issues the corresponding CXL.cache or CXL.mem transactions through the device request path. Communication flows through memory-mapped queues visible to both processor types.

Memory allocation in CXL memory uses a two-level allocator optimized for pointer-chasing access patterns. The first level maintains per-thread free lists for small allocations (up to 4KB), eliminating synchronization overhead. The second level manages large allocations from a global pool protected by a readers-writer lock. Both levels align allocations to cache line boundaries (64 bytes).

Thread management on the Vortex cores follows a work-stealing model that balances load across available warps. Each Vortex core maintains a local work queue of pending prefetch requests, processing them in approximate priority order. Warp scheduling uses a round-robin policy with priority boost for requests approaching their deadline.

\subsection{Memory Management and Coherence}

\sys does not implement a replacement coherence protocol in software. CXL's DCOH and the host coherence fabric perform the actual cache-line state transitions. What \sys maintains is a compiler/runtime shadow directory of \emph{requested} state: which lines have outstanding CS/CO/NC-P intent, which futures publish data to the host, and which Type~2 regions are currently treated as host-biased or device-biased. This metadata is used for legality checking, conflict avoidance, and policy selection; it is not the source of architectural coherence.

The shadow directory is partitioned across Vortex cores, with each core maintaining entries for a subset of CXL memory addresses. Directory lookups add 2--3 cycles when the entry is cached, or approximately 50ns when accessing CXL-resident entries. The directory fits entirely in Vortex on-chip memory for working sets up to 64GB. If a host-visible region is switched to device-bias mode, the runtime inserts an explicit phase boundary and rejects concurrent host writes until the region is returned to host-bias mode.




\section{Evaluation}
\label{sec:eval}

We evaluate \sys with three questions:
\begin{itemize}[leftmargin=*]
    \item \textbf{Q1} Does \sys improve end-to-end performance?
    \item \textbf{Q2} What aspects of \sys{}'s design contribute to improved performance?
    \item \textbf{Q3} What is the overhead of \sys{}'s techniques?
    \item \textbf{Q4} How does \sys{}'s performance compare to related work?
\end{itemize}

\paragraph{Experimental Setup} Our evaluation uses a heterogeneous CXL platform consisting of an Intel Xeon 6 6787p host with 256\,GB DDR5-8000 and a Type~2 Intel Agilex 7 FPGA implementing Vortex RISC-V SIMT cores.  We execute each benchmark with its data placed in the CXL-attached memory region and average the results of 10 trials in all reported numbers.

\begin{table}[t]
\caption{Workload characteristics. The data size reports the evaluated input or working-set size used in our experiments, rather than the benchmark-suite maximum.}
\resizebox{\columnwidth}{!}{%
\label{tab:workload_char}
\footnotesize
\begin{tabular}{@{}llll@{}}
\toprule
\textbf{Benchmark} & \textbf{Workload Type} & \textbf{Memory Bottleneck} & \textbf{Evaluated Data Size} \\
\midrule
GAPBS   & Graph processing       & Latency   & 240 MB \\
MCF     & Network optimization   & Latency   & 345 MB \\
Spatter & Irregular memory access & Latency   & 1 GB \\
NPB     & Scientific computing   & Bandwidth & 12.8 GB \\
\bottomrule
\end{tabular}%
}
\end{table}

\paragraph{Workloads}  We evaluate \sys on four benchmark suites (see \cref{tab:workload_char}).  GAPBS~\cite{gapbs} is a graph processing benchmark suite; we configure the workloads to execute over random graphs with $2^{20}$ vertices.  The GAPBS workloads perform pointer chasing operations; their performance is bounded by memory latency.  MCF performs a vehicle scheduling planning problem by modeling the problem as a graph and using a network-simplex algorithm.  MCF also exhibits pointer chasing behavior and is consequently bounded by memory latency.  Spatter is a benchmarking suite for gather and scatter accesses, which are common in high-performance computing~\cite{spatter_memsys20}.  Spatter consists of 3 workloads, we configure it to use a 1GiB dataset.  Spatter is bounded by memory latency because gather and scatter accesses are irregular memory access patterns for which hardware prefetchers perform poorly.  Lastly, we use the NAS parallel benchmark (NPB) suite~\cite{nas_benchmarks}, which is a collection of 8 scientific computing workloads derived from computational fluid dynamics.  The NPB workloads are bounded by memory bandwidth because they perform sequential memory accesses.

\begin{figure*}[t]
\centering
\caption{End-to-end speedup of \sys across benchmarks. Speedup is computed as host execution time divided by heterogeneous execution time with compiler-orchestrated overlap. The figure uses direct baseline labels, a two-row layout, host runtime annotations, and a geometric-mean group on the right.}
\label{fig:overall_speedup}
\includegraphics[width=\linewidth]{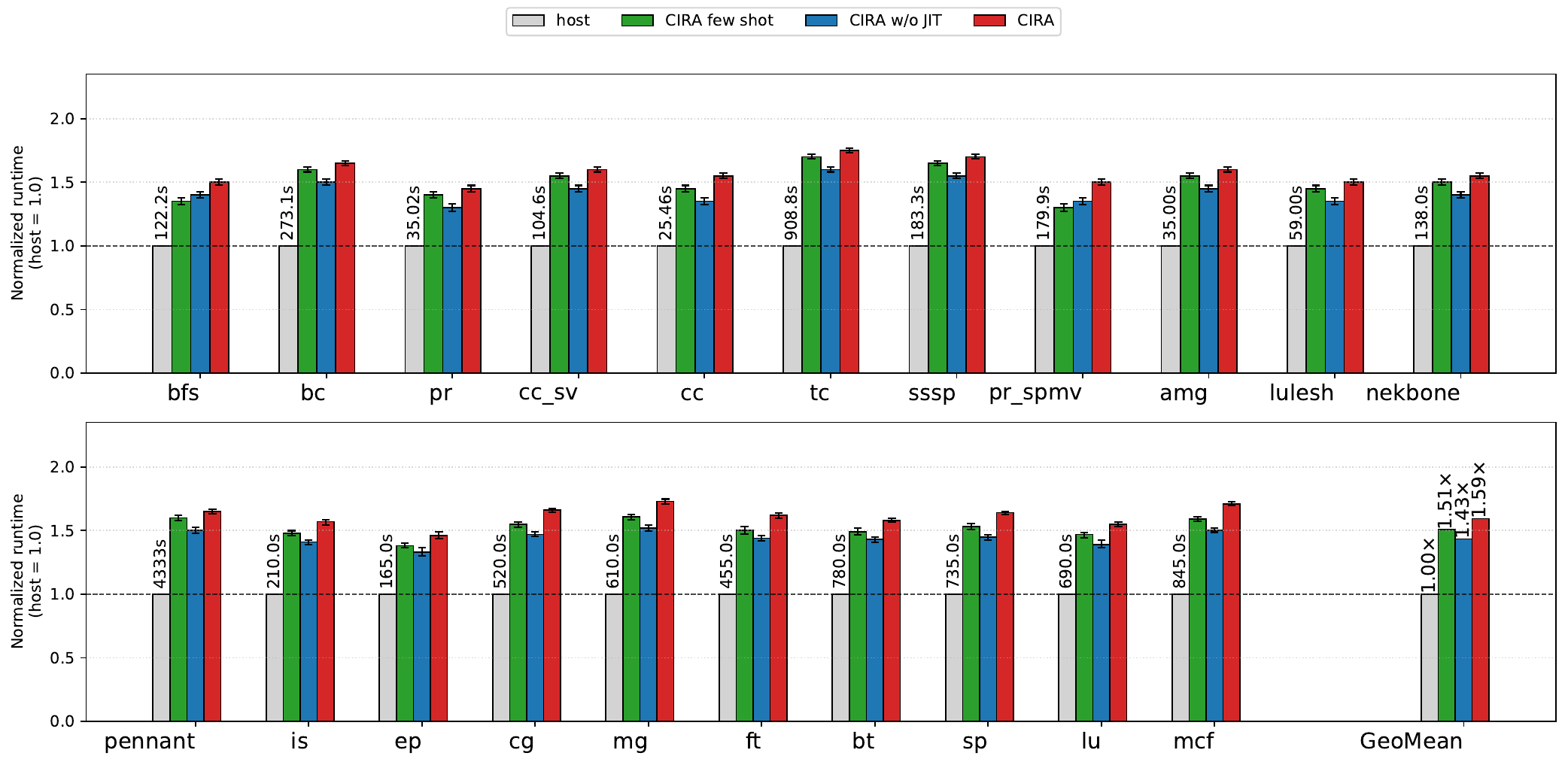}
\end{figure*}

\paragraph{Baselines}
We compare \sys against three configurations that isolate different aspects of the design: (1) \emph{Host}, where execution runs entirely on the host with no heterogeneous offload. (2) \emph{CIRA few shot}, which only places CXL load instructions into the \region; note: CIRA few shot allows load instructions that are direct, indirect, or doubly indirect (i.e., \texttt{*ptr}, \texttt{*A[i]}, and \texttt{*A[B[i]]}), but no additional computation (e.g., no control-flow).  Finally, (3) \emph{CIRA w/o JIT}, which uses the \sys compiler, but does not use the \sys JIT. 

\subsection{Application Performance}

We execute each of the workloads on each of the baselines; \cref{fig:overall_speedup} summarizes the results. \sys achieves a speedup of  \sysSpeedupLow--\sysSpeedupHigh$\times$ (\sysSpeedupGeomean$\times$ geometric mean).  The strongest gains arise in workloads whose critical path is dominated by dependent remote accesses, especially graph analytics and pointer-intensive optimization kernels, such as tc and sssp from GAPBS and mg from NPB.

We conclude that \sys improves end-to-end application performance.

\subsection{Understanding \sys{}'s Performance Across Configurations}

We answer Q2 by analyzing the performance gap between prefetch only, \sys w/o JIT and \sys.  At a high-level, we observe that \sys outperforms each of the other configurations, suggesting that its overall design contribute to an advantage compared to incomplete approaches. 

However, the intermediate results are surprising. First, CIRA few shot is quite effective on its own: it achieves a speedup of between 1.3--1.7$\times$ (1.51$\times$ geometric mean). The speedup occurs because CIRA few shot is capable of offloading fairly complex operations, which are evidently expensive enough to amortize the overheads of \sys{}'s software-managed asynchrony. Note that the \regions produced by CIRA few shot are significantly larger than the asynchronous operations of prior work for commodity systems; e.g., both hardware and software prefetchers only offload direct memory accesses.

Also interestingly, \sys{} w/o JIT performs worse than CIRA few shot. It achieves a speedup of between 1.3--1.6$\times$ (1.43$\times$ geometric mean). \sys{} w/o JIT places too many operations in its \regions, so too much of the code executes on the lower-performance \sys{} hardware accelerator, causing performance regressions relative to the more selective CIRA few shot mode. This performance gap demonstrates the importance of constructing \regions carefully.

In conclusion, we conclude that \sys{}'s performance comes in large part from the ability to offload complex load operations (i.e., CIRA few shot), but that reaching the highest benefit requires both creating and optimizing complex \regions.

\subsection{Understanding \sys{}'s Performance through Case Studies}

We further elaborate on Q2 by analyzing \sys{}'s performance on two candidate applications.

\subsubsection{GAPBS}
\begin{table}[t]
\caption{CXL vs. DRAM comparison for GAPBS. DRAM binds data to local DDR memory; CXL binds the same kernels to the CXL memory node while preserving CPU placement. Slowdown is relative to the DRAM run.}
\label{tab:gapbs_vtune_cxl}
\resizebox{\columnwidth}{!}{%
\begin{tabular}{@{}llll@{}}
\toprule
\textbf{Kernel} & \textbf{DRAM Stalls(Cyc)} & \textbf{CXL Stalls(Cyc)} & \textbf{CIRA Stalls(Cyc)} \\
\midrule
BFS      & 1.532B & 5.341B & 3.229B \\
BC       & 1.214B & 5.612B & 2.341B \\
PR       & 1.342B & 7.341B & 4.315B \\
CC\_SV   & 1.692B & 6.291B & 4.313B \\
PR\_SPMV & 1.892B & 7.401B & 4.327B \\
\bottomrule
\end{tabular}%
}
\end{table}

We use VTune to study the architectural effects of using \sys on a system with CXL memory.  We profiled each GAPBS kernel three times: once with data placed in local DRAM, once with data placed on CXL, and once with data placed in CXL using \sys{} to accelerate accesses.  We count the number of stalls that the program incurs throughout its execution in each of the configurations.

Table~\ref{tab:gapbs_vtune_cxl} provides the results. We observe that using CXL memory adds a considerable number of stalls to the DRAM configuration---the average number of stalls grows from 1.53B to 6.40B cycles, or 4.17$\times$. However, \sys reduces many of the stalls from the base CXL approach: the average number of stalls drops from 6.40B to 3.71B cycles, a 1.73$\times$ reduction compared to native CXL.


\subsubsection{MCF}

\begin{table}[t]
\caption{MCF per-\region profiling and optimization. Vortex prefetching reduces effective memory stall cycles by hiding pointer-chasing latency (\gainPerStep ns per chain step).}
\label{tab:mcf_detail}
\resizebox{\columnwidth}{!}{%
\begin{tabular}{@{}lrrrrr@{}}
\toprule
\textbf{Hotspot Functions} & \textbf{CXL Stalls} & \textbf{CIRA Stalls} & \textbf{CXL Cache} & \textbf{CIRA Cache} \\
 & \textbf{(cycles)} & \textbf{(cycles)} & \textbf{Hit \%}  & \textbf{Hit \%}\\ \midrule
\texttt{pricing\_kernel} & 32034 & 15332 & 0.25 & 0.49 \\
\texttt{price\_out\_impl} & 3256 & 5541 &  0.81 & 0.51 \\
\midrule
& 35290 & 20873\\
\bottomrule
\end{tabular}%
}
\end{table}

Next, we analyze the performance of MCF on \sys as well as when deployed natively on CXL.  We use vTune to measure the number of stalls that occur in both configurations on MCF's two hotspot functions (i.e., the functions that contribute most to CPU time).  We also calculate the cache hit \% of the host CPU when deployed in the CXL configuration (i.e., CXL Cache Hit \%) and on the \sys configuration (i.e., \sys Cache Hit \%).  \Cref{tab:mcf_detail} provides the results. 

We observe that \sys reduces the number of stalls cumulatively across the hotspot functions by 1.69$\times$.  The overall stall reduction is largely attributable to the dominant hotspot, \texttt{pricing\_kernel}: CIRA increases its cache hit rate from 25\% to 49\%, cutting stall cycles by 52.1\%. Although \texttt{price\_out\_impl} sees higher stalls under CIRA, the reduction in \texttt{pricing\_kernel} dominates, lowering total stalls from 35,290 to 20,873 cycles.  One interesting property is that the stall reduction is non-uniform: \sys actually incurs more stalls in \texttt{price\_out\_impl} compared to CXL.  

\subsection{Overheads}
\label{sec:overhead}

This subsection answers Q3 by measuring the fixed costs that \sys must amortize. We isolate two sources of overhead: (1) communication and memory-management overhead from writing descriptors, polling completion state, and installing cache lines; and (2) device compute overhead from loading and dispatching Vortex-side functions. The microbenchmark runs the same shared-buffer protocol used by the runtime on the Agilex 7 platform and reports per-operation latency in Table~\ref{tab:sync_overhead}.

The results suggest that the core primitives that \sys{} uses to communicate across host and device fall into two general sets.  First, some operations require roundtrip communication from the host or device to the Device Coherence Engine, such as \texttt{cira\_doorbell\_wait} poll, \texttt{cira\_doorbell\_ring}, and a cold Vortex \region load (i.e., a \region load that the Vortex core has not previously completed).  These operations take around .5\textmu s.  Second, some operations require roundtrip communication from the host or device to the Device Coherence Engine and also that the other endpoint also have round-trip communication with the Device Coherence Engine.  Such operations, including Host to Vortex BRAM reads and writes as well as invocations of \texttt{cira\_hoist\_exec}, take around 1\textmu s.

\begin{table}[t]
\caption{Runtime synchronization overhead breakdown (nanoseconds per operation). Measured on Intel Agilex 7 FPGA at \vortexFreqMhz MHz.}
\label{tab:sync_overhead}
\resizebox{\columnwidth}{!}{%
\begin{tabular}{@{}lr@{}}
\toprule
\textbf{Operation} & \textbf{Latency (ns)} \\ \midrule
Host write to Vortex BRAM & 1148 \\
Host read from Vortex BRAM & 1052 \\
Host \texttt{cira\_doorbell\_wait} poll latency & 510 \\
Vortex \texttt{cira\_doorbell\_ring} latency & 480 \\
Cold Vortex \region load  & 512 \\
Average Vortex \region load & 45 \\
Average \texttt{cira\_hoist\_exec} latency & \syncOverhead \\
\bottomrule
\end{tabular}
}
\end{table}

\subsection{Comparison To Prior State-of-the-Art}
\label{sec:gem5_amu}
  \begin{table}[t]
  \caption{gem5 comparison against AMU and Vanilla CXL. Results are normalized to Vanilla CXL. AMU improves host-side memory-level parallelism, while \sys additionally moves
  address-generation and cache-state orchestration to the CXL-attached device.}
  \resizebox{\linewidth}{!}{%
  \label{tab:gem5_amu_comparison}
  \centering
  \begin{tabular}{@{}lrrrr@{}}
  \hline
  \textbf{Benchmark} &
  \textbf{CXL Latency(ms)} &
  \textbf{AMU} &
  \textbf{CIRA} &
  \textbf{CIRA / AMU} \\
  \hline
  BFS & 220 & 0.82$\times$ & 2.86$\times$ & 3.49$\times$ \\
  BC &  293& 0.81$\times$ & 3.72$\times$ & 4.58$\times$ \\
  PR & 118 & 0.67$\times$ &  1.50$\times$ & 2.25$\times$ \\
  CC\_SV & 199 & 0.77$\times$ & 2.55$\times$ & 3.38$\times$ \\
  \hline
  Geomean & 197 & 0.76$\times$ & 2.52$\times$ & 3.30$\times$ \\
  \hline
  \end{tabular}%
  }
  \end{table}

We compare \sys to an alternative approach for hiding CXL memory latency, AMU\cite{wang2022async}.  AMU represents a different point in the CXL latency-mitigation design space. Rather than moving address-generation logic to a near-memory device, AMU adds asynchronous memory-access support inside the out-of-order CPU core.   Its AMI instructions separate request issue from response consumption, and the AMU tracks many outstanding requests using scratchpad and metadata storage carved from the L2 cache. This design increases memory-level parallelism within the host core, but it does not provide near-memory execution, device-side traversal.  Moreover, AMU requires changes to the host CPU to add the AMU logic.  I.e., AMU does not provide \sys{}'s commodity support.

Since AMU requires host CPU changes, we evaluate it using a cycle accurate simulator, Gem5~\cite{wang2026cohet,gem5}, and also create a \sys port to Gem5.  We run four benchmarks from GAPBS and measure the latency for both configurations as well as when using a CXL configuration of gem5. We model three configurations.  \emph{Vanilla CXL} uses the same host core but places benchmark data in the modeled CXL memory node. \emph{AMU} extends the host core with asynchronous memory requests and response polling, following the AMU execution model, but leaves all address generation and dependence traversal on the host. \emph{CIRA} models the same CXL latency as Vanilla CXL, but offloads the selected memory-side slice to a near-memory execution engine and applies the same \region-formation and CXL.cache policy-selection rules used in the hardware evaluation. All results are normalized to Vanilla CXL, so values greater than $1.0\times$ indicate improvement over an unmodified CXL-attached-memory execution.  \Cref{tab:gem5_amu_comparison} presents the results. 

We observe that AMU causes a slowdown when compared to vanilla CXL.  The key issue is that AMU's scratchpad and L2 cache compete for capacity, so the system observes a larger number of cache misses compared to vanilla CXL.  It is likely that AMU would perform well given the perfect configuration of L2 cache size and scratchpad size.  CIRA outperforms AMU.  CIRA's approach enables near-memory cores to run ahead of the host and prepare future cache lines before the host reaches the corresponding instruction. In this way, CIRA exploits more asynchronous operations than AMU.

\section{Related Work}

\paragraph{Far Memory and Disaggregated Memory Systems.}
A large body of systems established two central lessons. First, transparent page-based mechanisms are often too coarse for irregular or pointer-heavy codes, where the dominant cost comes from repeated dependent misses rather than from poor bulk page placement~\cite{ruan2020aifm,maruf2020leap,tauro2024trackfm}. Second, software-aware mechanisms become more effective as they move closer to the program's dependence structure, data layout, and access semantics~\cite{ruan2020aifm,wang2022memliner,tauro2024trackfm,guo2023mira}. Our work inherits this insight but targets a different hardware setting. Prior far-memory systems were primarily designed around RDMA-style remote memory, where data movement occurs through object- or page-granularity software paths and remote memory is outside the processor's coherent cache hierarchy~\cite{ruan2020aifm,maruf2020leap}. In contrast, CXL introduces a cache-coherent memory interface with substantially tighter integration between host processors and attached memory or accelerator devices~\cite{cxl30spec,cxlintro2024}. That shift changes the optimization target: instead of focusing only on paging, prefetching, or placement, a compiler can reason about overlap between host execution and memory-side orchestration in a shared-coherence setting. In this sense, \sys is closer in spirit to compiler-guided far-memory systems such as TrackFM than to purely OS-level tiering, but extends the idea to heterogeneous host--device execution over CXL-attached memory~\cite{tauro2024trackfm,cxl30spec}.

\paragraph{CXL Memory Expansion, Tiering, and Pooling.}
Recent CXL work has largely focused on memory expansion, tiering, topology effects, and pooling. The CXL standard itself defines coherent protocols for memory expansion and sharing, including the mechanisms that make device-attached memory visible as part of a coherent system memory hierarchy~\cite{cxl30spec}. Several recent studies have characterized the latency and bandwidth behavior of real CXL memory devices and topologies, showing that CXL memory is usable but substantially slower than local DRAM, with performance depending strongly on placement, topology, and access pattern~\cite{sun2024characteristic,w2024topology}. Other work has explored how operating systems and virtualized environments should manage CXL-attached memory through tiering and migration, for example via transparent page placement or transactional migration across memory tiers~\cite{maruf2023tpp,xiang2024nomad,zhong2024managing}. There is also active debate about the system-level value of CXL memory pooling, with some studies arguing that pooling may be less beneficial in practice than often assumed once cost, complexity, and workload structure are accounted for~\cite{levis2023caseagainst}.


\paragraph{Processing-in-Memory and Near-Data Execution.}
Processing-in-memory (PIM) and near-data processing (NDP) have long pursued the goal of reducing data movement by executing logic close to memory. Recent surveys show that this space spans analog in-memory compute, logic-layer NDP in 3D-stacked memory, and programmable near-memory cores, each with different tradeoffs in programmability, generality, and deployment complexity~\cite{asifuzzaman2023pimsurvey}. Real hardware results from UPMEM have been particularly important in grounding the discussion: Nider et al.\ evaluated a commercial PIM system in an off-the-shelf server and showed that while near-memory execution can accelerate memory-bound kernels, performance depends heavily on workload structure, available parallelism, and the partitioning of work between host and memory-side processors~\cite{nider2021upmem}. Gomez-Luna et al.\ likewise analyzed real PIM hardware across a broad benchmark suite and emphasized that memory-boundness alone is not sufficient; performance also depends on irregularity, synchronization cost, data distribution, and the overheads of coordinating host and PIM execution~\cite{gomezluna2021analysis}.


\paragraph{Fixed-Function and Programmable Data-Access Accelerators.}
A more targeted line of work accelerates memory access itself rather than general computation. The closest recent example is DX100, a programmable data-access accelerator for indirect memory accesses. DX100 offloads address calculation and bulk data-access orchestration to a shared accelerator and provides a compact ISA plus compiler support for transforming legacy code to use the device~\cite{khadem2025dx100}. This is an important step toward treating indirection as a first-class acceleration target. However, DX100 is designed around a dedicated access accelerator integrated into the system's memory path, with an instruction set specialized for indirect access patterns and a hardware design aimed at improving DRAM-level bandwidth utilization through reordering and coalescing~\cite{khadem2025dx100}. By contrast, \sys does not assume a fixed-function access engine or a predefined access ISA. Instead, it generates device-side code from the program itself and executes that code on near-memory programmable cores attached through CXL. This makes \sys less specialized than DX100 but also more flexible: the outlined device region can incorporate workload-specific dependence traversal, prefetch triggering, and cache-management behavior that need not fit a small fixed instruction vocabulary. More broadly, \sys differs from both classic prefetchers and access accelerators in that its main abstraction is heterogeneous overlap between host computation and memory-side orchestration, not just earlier issuance of memory requests.

\section{Conclusion}

The CXL era necessitates a rethinking of the hardware-software contract. The assumption that memory is a passive, uniform resource is no longer tenable in a world of high-latency, disaggregated media with embedded compute capabilities and programmable coherence-state behavior. Our VTune-driven workload characterization demonstrates that memory bottlenecks are both pervasive and structurally diverse---with hotspots that shift by query type, access pattern, and cache hierarchy level---ruling out fixed-ISA solutions and motivating compiler-driven approaches. \sys demonstrates that by exposing the heterogeneity of modern CXL memory expanders to the compiler, generating flexible, asynchronous offload strategies with true computation--communication overlap, we can fundamentally alter the performance characteristics of the system. Through the \cira intermediate representation and the coordinated execution of x86 and Vortex cores, \sys effectively hides the ``CXL Tax,'' turning the latency liability into a throughput asset. As datacenters continue to disaggregate, compiler-driven techniques like \sys will be essential to bridging the gap between capacity and performance.

\bibliographystyle{ACM-Reference-Format}
\bibliography{bibliography}

\end{document}